\documentclass[12pt]{iopart}

\usepackage{iopams}
\usepackage{graphicx}

\usepackage{xcolor}
\expandafter\let\csname equation*\endcsname\relax
\expandafter\let\csname endequation*\endcsname\relax
\usepackage{amsmath}
\usepackage{bm}
\usepackage{braket}

\newcommand{\pdiff}[2]{\frac{\partial #1}{\partial #2}}

\newcommand{\new}{\nonumber\\}
\newcommand{\bx}{\bm{x}}

\newcommand{\be}{\bm{e}}

\newcommand{\abs}[1]{\left|#1\right|}
\newcommand{\ave}[1]{\left\langle #1\right\rangle}

\newcommand{\pr}{{\rm PR}}

\begin{document}

\title{Bose-Einstein-Like condensation of deformed random matrix: A replica approach}

\author{Harukuni Ikeda$^{1}$}

\address{
$^1$Department of Physics, Gakushuin University, 1-5-1 Mejiro, Toshima-ku, Tokyo 171-8588, Japan
}
\ead{harukuni.ikeda@gakushuin.ac.jp}
\vspace{10pt}
\begin{indented}
\item[]\today 
\end{indented}

\begin{abstract}
In this work, we investigate a symmetric deformed random matrix, which
is obtained by perturbing the diagonal elements of the Wigner
matrix. The eigenvector $\bx_{\rm min}$ of the minimal eigenvalue
$\lambda_{\rm min}$ of the deformed random matrix tends to condensate at
a single site. In certain types of perturbations and in the limit of the
large components, this condensation becomes a sharp phase transition,
the mechanism of which can be identified with the Bose-Einstein
condensation in a mathematical level. We study this Bose-Einstein like
condensation phenomenon by means of the replica method. We first derive
a formula to calculate the minimal eigenvalue and the statistical
properties of $\bx_{\rm min}$. Then, we apply the formula for two
solvable cases: when the distribution of the perturbation has the double
peak, and when it has a continuous distribution.  For the double peak,
we find that at the transition point, the participation ratio changes
discontinuously from a finite value to zero.  On the contrary, in the
case of a continuous distribution, the participation ratio goes to zero
either continuously or discontinuously, depending on the distribution.
\end{abstract}


\section{Introduction}

In this manuscript, we study the eigenvector $\bx_{\rm min}$ of the
minimal eigenvalue $\lambda_{\rm min}$ of the deformed Wigner matrix,
where the $i$-th diagonal element of the Wigner matrix is perturbed by a
constant $h_i$~\cite{rosenzweig1960,kravtsov2015random,
capitaine2016spectrum,facoetti2016non,lee2016extremal}. When
$\abs{h_i}\ll 1$, all components of $\bx_{\rm min}$ have the same order
of magnitude, as in the case of the original Wigner
matrix~\cite{livan2018introduction}. On the contrary, when $\abs{h_i}\gg
1$, $\bx_{\rm min}$ tends to condensate at the site with the smallest
$h_i$~\cite{lee2016extremal}. For some specific distributions of $h_i$,
the condensation becomes a sharp phase transition in the limit of the large
number of components~\cite{kravtsov2015random}. Interestingly, this
condensation transition has a similar mathematical structure of that of
the Bose-Einstein
condensation~\cite{aspelmeier2004,kravtsov2015random,franz2022delocalization,franz2022linear}.

The deformed random matrix has been used to understand complex atomic
spectra~\cite{rosenzweig1960}, Anderson
Localization~\cite{kravtsov2015random}, principal component
analysis~\cite{perry2018optimality}, and so
on~\cite{krajen2021}. Recently the model has gained renewed interest as
a toy model to describe the vibrational properties of amorphous 
solids~\cite{franz2022delocalization,franz2022linear,rainone2021mean,bouchbinder2021low}.
Several numerical studies uncovered that in addition to the usual phonon
modes, there appear many quasi-localized modes in low-frequency
vibrational density of states of amorphous
solids~\cite{lerner2016,mizuno2017continuum,angelani2018probing,wang2019low}.
In particular, the participation ratio of the quasi-localized mode of
the lowest frequency is inversely proportional to the system size,
meaning that the eigenvector of the minimal eigenvalue of the Hessian of
an amorphous solid is localized~\cite{lerner2016}. The result
contradicts a mean-field theory of the glass transition, where a Hessian
of an amorphous solid is approximated by a dense random matrix, whose
eigenvectors are
extended~\cite{castellani2005spin,franz2015universal}. To reconcile this
discrepancy, Rainone {\it et
al.}~\cite{rainone2021mean,bouchbinder2021low,folena2022marginal}
recently introduced a mean-field model whose effective Hessian in the RS
phase can be considered as a deformed random matrix. The model 
exhibits the localization transition at which the eigenvector of the
minimal eigenvalue is localized. Thus it correctly reproduces the
localized property of amorphous solids. More recently, Franz {\it et
al.}  studied a fully-connected vector spin-glass model and found
similar localization of the eigenvector of the lowest
frequency~\cite{franz2022delocalization,franz2022linear}.  They also
pointed out that this localization is caused by a Bose-Einstein (like)
condensation~\cite{franz2022delocalization,franz2022linear}.

Motivated by those recent developments of disordered systems, here we
investigate a replica method to describe the Bose-Einstein like
condensation of the minimal eigenvector of the deformed random
matrix. The replica method is a powerful tool to treat disordered
systems such as spin-glass~\cite{mezard1987spin}, amorphous solids, and
granular materials~\cite{parisi2020theory}. This is also true in the
field of random matrices~\cite{livan2018introduction}. A seminar work
has been done by Edwards and Jones in Ref~\cite{edwards1976eigenvalue}.
They studied a symmetric random matrix in which each element follows a
Gaussian distribution of zero mean and fixed variance. By using the
replica method, they showed that the eigenvalue distribution of the
matrix converges to the well-known Wigner semicircle
distribution~\cite{livan2018introduction}. Later, the replica method was
also applied to calculate the eigenvalue distribution of an asymmetric
random matrix~\cite{sommers1988}, symmetric sparse random
matrix~\cite{rodgers1988,semerjian2002sparse,nagao2007spectral,kuhn2008spectra},
and so on. We here show that the replica method is also useful for the
analysis of the lowest eigenmode of the deformed random matrix.

The structure of the paper is as follows. In Sec.~\ref{185920_1Aug22},
we describe the model. In Sec.~\ref{185931_1Aug22}, we describe how to
calculate the minimal eigenvalue and eigenvector by using the replica
method. In Sec.~\ref{190442_1Aug22}, we present the results. In
Sec.~\ref{190505_1Aug22}, we conclude the work.

\section{Model}
\label{185920_1Aug22} We consider a $N\times N$ symmetric matrix whose
$ij$ component is written as
\begin{align}
 W_{ij} = J_{ij} + h_i\delta_{ij}.\label{080311_2Aug22}
\end{align}
Here $J_{ij}=J_{ji}$ is a i.i.d random variable 
following a Gaussian distribution
\begin{align}
P(J_{ij}) = \sqrt{\frac{N}{2\pi}}e^{-\frac{NJ_{ij}^2}{2}},\label{192347_1Aug22}
\end{align}
and $\{h_i\}_{i=1,\dots, N}$ are constants. Unfortunately, our present
method does not work for general values of $h_i$'s. We restrict our
analysis for a specific case~\cite{urbani2022field}:
\begin{align}
 h_i =
\begin{cases}
h_1 & (i = 1,\dots, N/M),\\
\vdots \\
h_k & (i = kN/M+1,\dots, (k+1)N/M),\\
\vdots\\
h_M & (i = N-N/M+1,\dots, N).\\
\end{cases}\label{140017_3Aug22}
\end{align}
By setting $h_i$ this way, we can define an overlap $q_k$ corresponding
to each $h_k$, which quantifies how much the eigenvector is
condensed/localized to the sites perturbed by $h_k$. At the end of the
calculation, we take the $M\to\infty$ limit, but even there we require
that $N/M$ goes to infinity. To be more specific, we first take the
thermodynamic limit $N\to\infty$ and then take the limit $M\to \infty$.


\section{Theory}
\label{185931_1Aug22}
\subsection{Interaction potential and ground state}

Here we use the method developed by Kabashima and
Takahashi~\cite{kabashima2012first}. To investigate the minimal
eigenvalue $\lambda_{\rm min}$ and corresponding vector $\bx_{\rm min}$
of $W$, we consider a system interacting with the following potential:
\begin{align}
H(\bx|J) &\equiv  \frac{\bx\cdot W\cdot \bx}{2}\new 
& = \frac{\bx\cdot J\cdot \bx}{2} + \frac{1}{2}\sum_{k=1}^M h_k\bx_k\cdot\bx_k,\label{001213_31Jul22}
\end{align}
where the $N$ dimensional vector $\bx=\{x_1,\dots, x_N\}$ denotes the
state variable. We also introduced the sub-vectors:
\begin{align}
\bx_k = \{x_i\}_{i= k\frac{N}{M}+1,\dots, (k+1)\frac{N}{M}}.
\end{align}
We impose that the state vector $\bx$ satisfies the spherical
constraint:
\begin{align}
\bx\cdot\bx = \sum_{k=1}^M \bx_k\cdot\bx_k = \sum_{i=1}^N x_i^2 = N.\label{012322_31Jul22}
\end{align}
When $h_i=0$, the model Eq.~(\ref{001213_31Jul22}) can be identified
with the $p=2$ spin spherical model, which has been fully investigated
before~\cite{sherrington1975,nieuwenhuizen1995,cugliandolo1995full}.

Under the above setup, it is easy to show that when $\bx=\bx_{\rm min}$,
we get the ground state energy~\cite{kabashima2012first,fyodorov2015large}:
\begin{align}
H_{\rm GS} = \frac{\bx_{\rm min}\cdot W\cdot \bx_{\rm min}}{2}=\frac{\lambda_{\rm min}}{2}N.\label{001006_31Jul22}
\end{align}
Therefore, the minimal eigenvalue is calculated as $\lambda_{\rm
min}=2H_{\rm GS}/N$.

\subsection{Replica method}
\label{190125_1Aug22}

To investigate the ground state, we introduce the partition
function~\cite{kabashima2012first}:
\begin{align}
Z(J) = \int d\bx \delta(\bx\cdot\bx -N) e^{-\beta H(\bx|J)}
\end{align}
and the free-energy
\begin{align}
 -\beta f = \frac{1}{N}\overline{\log Z(J)},\label{002304_31Jul22}
\end{align}
where $\beta=1/T$ denotes the inverse temperature, and the overline
denotes the average for the quenched randomness $J$.  The ground state
energy per particle is given by taking the zero temperature limit of the
free-energy
\begin{align}
e_{\rm GS} \equiv \frac{H_{\rm GS}}{N} = \lim_{T\to 0}f. 
\end{align}
Below we omit the subscript ${\rm GS}$ unless it causes confusion. To
perform the disordered average in Eq.~(\ref{002304_31Jul22}), we use the
replica trick~\cite{mezard1987spin}:
\begin{align}
 -\beta f = \lim_{n\to 0}\frac{\log\overline{Z(J)^n}}{nN},
\end{align}
where we have introduced the replicated partition function as follows:
\begin{align}
\overline{Z^n}
= \int \prod_{a=1}^n d\bx_a \delta(\bx_a\cdot\bx_a-N)
\overline{e^{-\beta \sum_{a=1}^n H(\bx_a|J)}}.
\end{align}
Since the distribution of $J_{ij}$ is a Gaussian
Eq.~(\ref{192347_1Aug22}), the quenched average can be taken
analytically~\cite{castellani2005spin}~\footnote{ Here and in subsequent calculations, we omit
constants and sub-leading terms that are not relevant to the final
result.}:
\begin{align}
 \overline{e^{-\beta \sum_{a=1}^n H(\bx_a|J)}}\sim
 \exp\left[\frac{N\beta^2}{4}\sum_{ab}Q^{ab} -\frac{\beta}{2M}\sum_{a=1}^n\sum_{k=1}^M h_kQ_k^{aa}\right],
\end{align}
where we have defined the overlaps as follows:
\begin{align}
&Q_k^{ab} \equiv \frac{\bx_k^a\cdot\bx_k^b}{N/M},\new
&Q_{ab} \equiv \frac{1}{M}\sum_{k=1}^M Q_k^{ab} = \frac{1}{N}\sum_{i=1}^N x_i^a x_i^b.
\end{align}
When we change the variable from $\{\bx_k^a\}_{a=1,\dots,n}$ to
$\{Q_k^{ab}\}_{a,b=1,\dots n}$, the following Jacobian
apepars~\cite{franz2017}:
\begin{align}
\prod_{a=1}^n\int d\bx_k^a
=  \prod_{a=1}^n
  \int d\bx_k^a \prod_{ab}
 \int dQ_k^{ab}\delta\left(\frac{N}{M}Q_k^{ab}-\bx_k^a\cdot\bx_k^b\right)
 \sim \prod_{ab}\int dQ_k^{ab}
 e^{\frac{N}{2M}\log \det Q_k}.
\end{align}
Summarizing the above results, we get
\begin{align}
\overline{Z^n}
 = \prod_{k,a,b}\int dQ_k^{ab} e^{NS(Q)},
\end{align}
where
\begin{align}
S(Q) &= \frac{1}{2M}\sum_{k=1}^M\log\det Q_k+ \frac{\beta^2}{4}\sum_{ab}\left(Q^{ab}\right)^2
 -\beta\frac{1}{2M}\sum_{a=1}^n \sum_{k=1}^M h_k Q_k^{aa}.
\end{align}
We should minimize $S(Q)$ with the spherical constraint 
\begin{align}
Q^{aa} = \frac{1}{M}\sum_{k=1}^M Q_k^{aa} = 1,\ a = 1,\dots, n.
\end{align}
To proceed the calculation, we assume the replica symmetric
Ansatz~\cite{mezard1987spin}:
\begin{align}
 Q_k^{ab} = \delta_{ab}q_k + (1-\delta_{ab})p_k.
\end{align}
Then, we get 
\begin{align}
S(Q)
&= \frac{1}{M}\sum_{k=1}^M\frac{1}{2}\left[\log(q_k+(n-1)p_k)+(n-1)\log(q_k-p_k)\right]\new 
&+ \frac{\beta^2}{4}\left(n+n(n-1)p^2\right)
-n\frac{\beta}{2M}\sum_{k=1}^M h_k q_k,
\end{align}
where
\begin{align}
 p = \frac{1}{M}\sum_{k=1}^M p_k.
\end{align}
Finally, by taking the $n\to 0$ limit, we get the free-energy
\begin{align}
-\beta f 
&= \lim_{n\to 0}\frac{\log \overline{Z^n}}{nN}
= \lim_{n\to 0}\frac{S(Q)}{n}\new 
&= \frac{1}{2M}\sum_{k=1}^M\left[\frac{p_k}{q_k-p_k}+\log(q_k-p_k)\right]
 + \frac{\beta^2}{4}(1-p^2)
 -\frac{\beta}{2M}\sum_{k=1}^M h_k q_k.\label{193441_1Aug22}
\end{align}

\subsection{Ground state energy}
To get the ground state energy, we should take the zero temperature
limit $T\to 0$. This is possible by using the harmonic approximation:
\begin{align}
T\chi_k = q_k-p_k,\label{193435_1Aug22}
\end{align}
which is validated at sufficiently low $T$~\cite{franz2017}.
Substituting Eq.~(\ref{193435_1Aug22}) into Eq.~(\ref{193441_1Aug22})
and taking $T\to 0$ limit, we get
\begin{align}
e = \lim_{T\to 0}f
=-\frac{1}{2M}\sum_{k=1}^M\frac{q_k}{\chi_k} 
 -\frac{\chi}{2}+\frac{1}{2M}\sum_{k=1}^M h_k q_k,
\label{084931_29Jul22}
\end{align}
where
\begin{align}
 \chi = \frac{1}{M}\sum_{k=1}^M \chi_k.
\end{align}

Now we minimize it for $\chi_k$ and $q_k$.  We first consider the saddle
point condition for $\chi_k$:
\begin{align}
\pdiff{e}{\chi_k} = \frac{q_k}{2M\chi_k^2} -\frac{1}{2M} = 0 \to \chi_k = \sqrt{q_k}.\label{091800_29Jul22}
\end{align}
Using this equation, one can eliminate $\chi_k$ from Eq.~(\ref{084931_29Jul22}):
\begin{align}
 e &= -\frac{1}{2M}\sum_{k=1}^M \sqrt{q_k} -\frac{1}{2M}\sum_{k=1}^M \sqrt{q_k}
 + \frac{1}{2M}\sum_{k=1}^M h_k q_k\new
 &=
 -\frac{1}{M}\sum_{k=1}^M \sqrt{q_k}+\frac{1}{2M}\sum_{k=1}^M h_k q_k.\label{141814_31Jul22}
\end{align}
Next, we should minimize $e$ w.r.t $q_k$ with the spherical constraint
$q=\sum_{k=1}^M q_k/M = 1$. To this purpose, we introduce the Lagrange
multiplier $\mu$:
\begin{align}
 e &= 
 -\frac{1}{M}\sum_{k=1}^M \sqrt{q_k}+\frac{1}{2M}\sum_{k=1}^M h_k q_k
 + \frac{\mu}{2M}\left(\sum_{k=1}^M q_k -M\right).
\end{align}
The saddle point condition for $q_k$ leads to 
\begin{align}
\pdiff{e}{q_k} = -\frac{1}{2M\sqrt{q_k}} + \frac{h_k+\mu}{2M} = 0
\to  \sqrt{q_k} = \frac{1}{\mu+h_k}.
\end{align}
Since $\sqrt{q_k}\geq 0$, $\mu$ should satisfy
\begin{align}
\mu +\min_k h_k\geq 0.\label{024515_31Jul22}
\end{align}
The Lagrange multiplier $\mu$ should be determined by the following
condition:
\begin{align}
 1 = \frac{1}{M}\sum_{k=1}^M q_k = \frac{1}{M}\sum_{k=1}^M\frac{1}{(\mu+h_k)^2}
 = \int_{-\infty}^\infty dh P(h)q(h),\label{203359_30Jul22}
\end{align}
where we have introduced the distribution of $h_k$ 
\begin{align}
 P(h) = \frac{1}{M}\sum_{k=1}^M \delta(h-h_k),
\end{align}
and the self-overlap of spins subjected to the external field $h$
\begin{align}
 q(h) = \frac{1}{(\mu+h)^2}.
\end{align}
Similar equations as Eq.~(\ref{203359_30Jul22}) have been previously
obtained for a sparse random matrix~\cite{kabashima2012first} and
deformed random
matrices~\cite{franz2022delocalization,franz2022linear,krajen2021}. Substituting
the above results into Eq.~(\ref{001006_31Jul22}), one can calculate
$\lambda_{\rm min}$ as follows:
\begin{align}
\lambda_{\rm min} &= \frac{2H_{\rm GS}}{N} = 2e\new 
 &= 2\int_{-\infty}^\infty dh P(h)\left[\frac{hq(h)}{2}-\sqrt{q(h)}\right]\new 
 &= 2\int_{-\infty}^\infty dh P(h)
 \left[
\frac{h}{2(h+\mu)^2}-\frac{1}{h+\mu}
 \right].\label{142721_31Jul22}
\end{align}

\section{Results}
\label{190442_1Aug22}

\subsection{Single delta peak}
We first check the result for a single delta peak:
\begin{align}
P(h) = \delta(h-\Delta),
\end{align}
which is tantamount to consider the matrix:
\begin{align}
W = J + \Delta I, 
\end{align}
where $I$ is the $N\times N$ identity matrix. The minimal eigenvalue of
this matrix is $\lambda_{\rm
min}=-2+\Delta$~\cite{livan2018introduction}. Below, we check if our
method can correctly reproduce this result.

The spherical constraint Eq.~(\ref{203359_30Jul22}) in this case is 
\begin{align}
 1 = \int_{-\infty}^\infty dhP(h)q(h) = q(\Delta) = \frac{1}{(\mu+\Delta)^2}.
\end{align}
Solving this equation for $\mu$, we get
\begin{align}
 \mu = 1-\Delta.
\end{align}
The minimal eigenvalue is calculated as 
\begin{align}
\lambda_{\rm min} =2e = 2\int_{-\infty}^\infty dh P(h)
 \left[
\frac{h}{2(h+\mu)^2}-\frac{1}{h+\mu}
 \right]
 = \frac{\Delta}{(\Delta+\mu)^2}-\frac{2}{\Delta+\mu} = \Delta-2.\label{160922_31Jul22}
\end{align}
The known result has been correctly reproduced.

\subsection{Binary distribution}
\begin{figure}[t]
\begin{center}
\includegraphics[width=10cm]{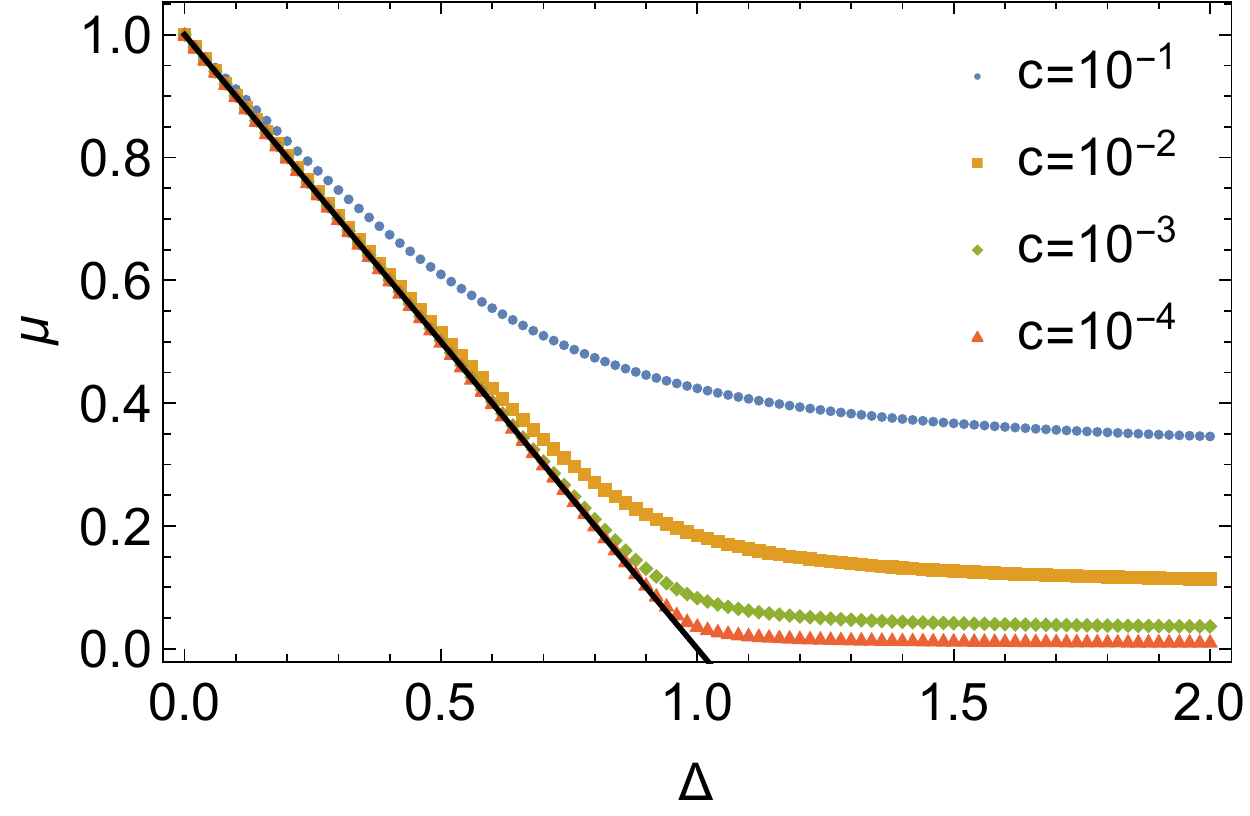} \caption{$\Delta$ dependence of
the Lagrange multiplier $\mu$ for the binary distribution.  Markers
denote the results for $c>0$, while the solid line denotes the result in
the limit $c\to 0$.}  \label{121345_31Jul22}
\end{center}
\end{figure}

\begin{figure}[t]
\begin{center}
\includegraphics[width=15cm]{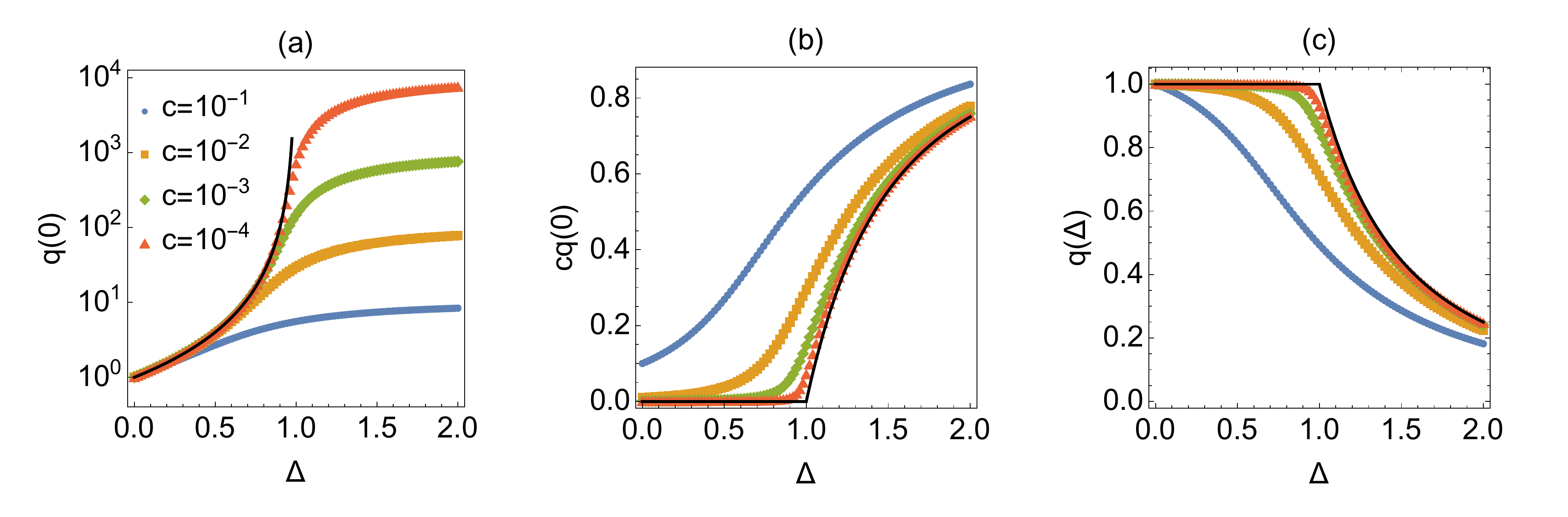}
\caption{$\Delta$ dependence of the overlaps for the binary distribution. 
Markers denote results for $c>0$,
 while the solid line denotes the result in the limit $c\to 0$.}
\label{131517_31Jul22}
\end{center}
\end{figure}

\begin{figure}[t]
\begin{center}
\includegraphics[width=10cm]{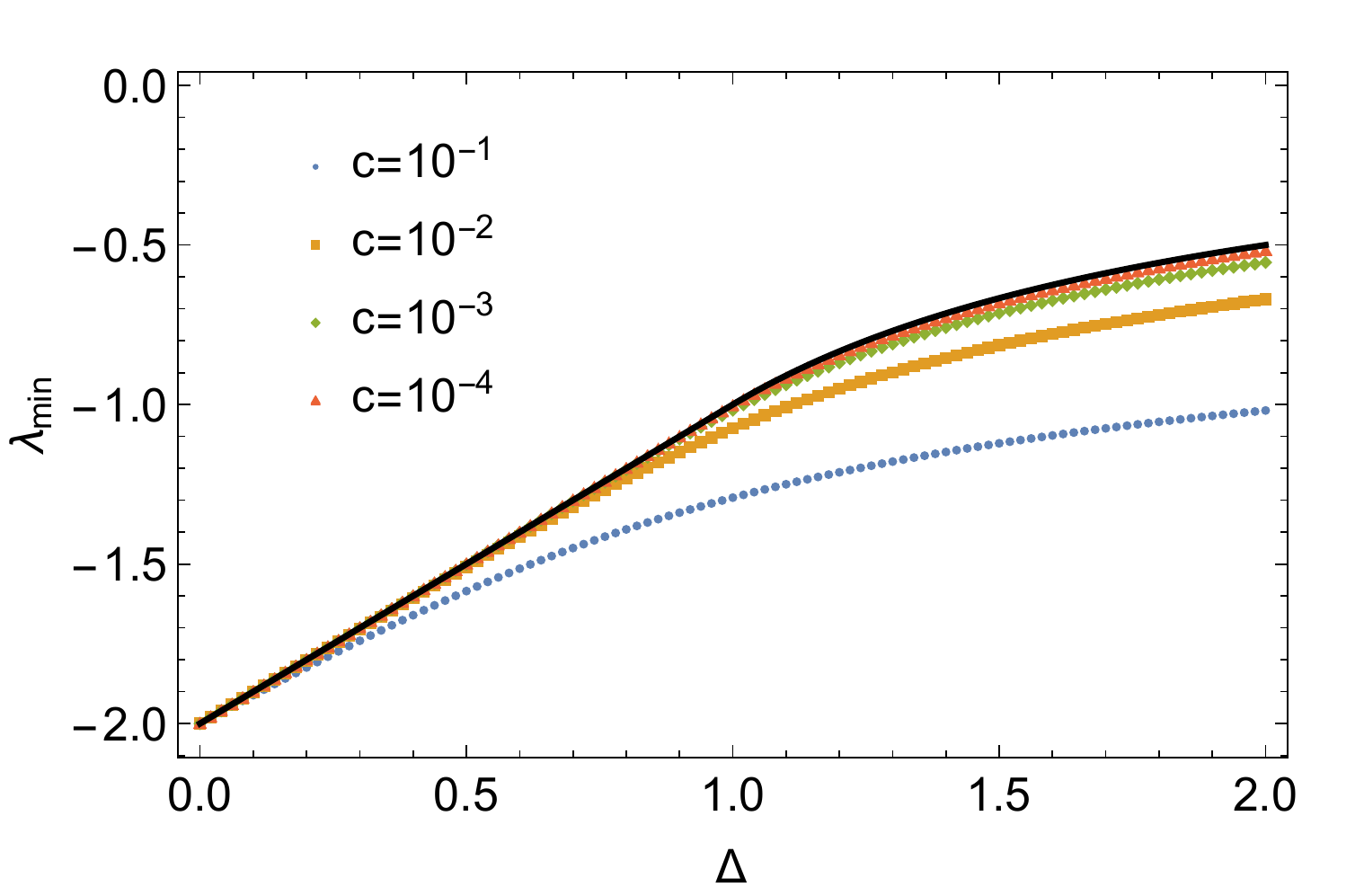} \caption{$\Delta$ dependence of
the minimal eigenvalue $\lambda_{\rm min}$ for the binary distribution.
Markers denote results for $c>0$, while the solid line denotes the
result in the limit $c\to 0$.}  \label{162016_31Jul22}
\end{center}
\end{figure}

Here we consider a simple binary distribution:
\begin{align}
 P(h) = c\delta(h) + (1-c)\delta(h-\Delta),\label{093028_31Jul22}
\end{align}
where $c\in [0,1]$ and $\Delta$ is a positive constant. Assuming the
distribution Eq.~(\ref{093028_31Jul22}) is tantamount to set the
external field in Eq.~(\ref{140017_3Aug22}) as
\begin{align}
h_i=
\begin{cases}
 0  & i=1,\dots, cN,\\
 \Delta & i = cN+1,\dots, N
 \end{cases}.
\label{binary}
\end{align}
Now the spherical constraint Eq.~(\ref{203359_30Jul22}) is written as follows
\begin{align}
1 = cq(0) + (1-c)q(\Delta),\label{024248_31Jul22}
\end{align}
where
\begin{align}
q(0) = \frac{1}{\mu^2},\hspace{5mm}
q(\Delta) = \frac{1}{(\mu+\Delta)^2}.\label{105758_31Jul22}
\end{align}
The Lagrange multiplier $\mu$ should be determined so as to satisfy
Eq.~(\ref{024248_31Jul22}). In Fig.~\ref{121345_31Jul22}, we plot $\mu$
for several $c$.
For later comparison with the result of the continuous distribution,
we are in particular interested in the limit $c\to
0$. A naive expectation is that Eq.~(\ref{024248_31Jul22}) in this limit
reduces to
\begin{align}
1 \approx q(\Delta) = \frac{1}{(\mu+\Delta)^2}.
\end{align}
Solving this equation, we get
\begin{align}
 \mu = 1-\Delta.\label{105807_31Jul22}
\end{align}
Eq.~(\ref{105807_31Jul22}) however implies that $\mu$ becomes negative
when $\Delta>1$, which is prohibited by Eq.~(\ref{024515_31Jul22}). What
was wrong?  What we missed is that when $\mu\sim 0$, the first term on
the right-hand side of Eq.~(\ref{024248_31Jul22}), $cq(0) = c/\mu^2$ can
no longer be ignored. Let we assume that this term takes a finite value
for $\Delta>1$, then from Eq.~(\ref{024248_31Jul22}), we get
\begin{align}
cq(0) = 1- \frac{1-c}{(\mu+\Delta)^2}\approx
 1- \frac{1}{\Delta^2}.\label{105812_31Jul22}
\end{align}
From Eqs.~(\ref{105758_31Jul22}), (\ref{105807_31Jul22}), and
(\ref{105812_31Jul22}), we can deduce the behavior of $\mu$, $q(0)$ and
$q(\Delta)$ in the limit $c\to 0$ as follows:
\begin{align}
& \mu = 
 \begin{cases}
  1-\Delta & (\Delta\leq 1)\\
  0        & (\Delta > 1)
 \end{cases},&
& q(0) =
 \begin{cases}
  1/(1-\Delta)^2 & (\Delta\leq 1)\\
  c^{-1}(1-1/\Delta^2)  & (\Delta > 1)
 \end{cases},&
&q(\Delta) =
  \begin{cases}
  1 & (\Delta\leq 1)\\
  1/\Delta^2 & (\Delta > 1)
 \end{cases}.\label{215423_1Aug22}
\end{align}
In Figs.~\ref{121345_31Jul22} and \ref{131517_31Jul22}, we plot $\mu$,
$q(0)$, and $cq(0)$ for several $c$ to show how these results converge
to Eqs.~(\ref{215423_1Aug22}) in the limit $c\to 0$.
From Eq.~(\ref{142721_31Jul22}),  ground state energy is calculated as
\begin{align}
 e = \int_{-\infty}^\infty dh P(h)
 \left[\frac{h}{2(h+\mu)^2}-\frac{1}{h+\mu} \right]
 =-\frac{c}{\mu} + (1-c)\left[\frac{\Delta}{2(\Delta+\mu)^2}-\frac{1}{\Delta+\mu} \right].
\end{align}
Substituting Eqs.~(\ref{215423_1Aug22}) into the above equation, we get
in the limit $c\to 0$
\begin{align}
\lambda_{\rm min} = 2e \to \begin{cases}
      \Delta-2 & (\Delta\leq 1),\\
      -1/\Delta & (\Delta>1).
     \end{cases}
\label{c0}
\end{align}
In Fig.~\ref{162016_31Jul22}, we plot this equation with the results of
finite $c$'s.

Now we discuss the degree of the localization. For this purpose, we
define the participation ratio:
\begin{align}
\pr \equiv  \frac{1}{N}\frac{(\sum_{i=1}^N \ave{x_i^2})^2}{\sum_{i=1}^N \ave{x_i^4}}
=\left[\frac{1}{N}\sum_{i=1}^N \ave{x_i^4}\right]^{-1},
 \label{101730_31Jul22}
\end{align}
where
\begin{align}
 \ave{O} \equiv \lim_{T\to 0}\frac{\int d\bx e^{-\beta H}O}{\int d\bx e^{-\beta H}}.
\end{align}
The partition ratio takes $\pr=O(1)$ when $\bx$ is extended, while
$\pr=0$ when $\bx$ is localized. To calculate the forth moment of $x_i$,
we assume that $x_i$ follows the normal distribution of zero mean and
variance $q(0)$ for $i\leq cN$ and variance $q(\Delta)$ for
$i>cN$~\cite{franz2022delocalization}.  Then, we get
\begin{align}
 \ave{x_i^4} \approx 3\ave{x_i}^2 = \begin{cases}
		3q(0)^2 & i = 1,\dots ,cN\\
		3q(\Delta)^2 & i = cN+1,\dots, N
	       \end{cases}.
\end{align}
In the limit $c\to 0$, Eq.~(\ref{101730_31Jul22}) reduces to
\begin{align}
\pr(\Delta) =  \frac{1}{3}\frac{1}{cq(0)^2+(1-c)q(\Delta)^2} 
\to 
\begin{cases}
1/3 & (\Delta\leq 1)\\
 0            & (\Delta > 1).
\end{cases}\label{141607_31Jul22}
\end{align}
Therefore, the eigenvector of the minimal eigenvalue is localized for
$\Delta>1$.
\begin{figure}[t]
\begin{center}
\includegraphics[width=10cm]{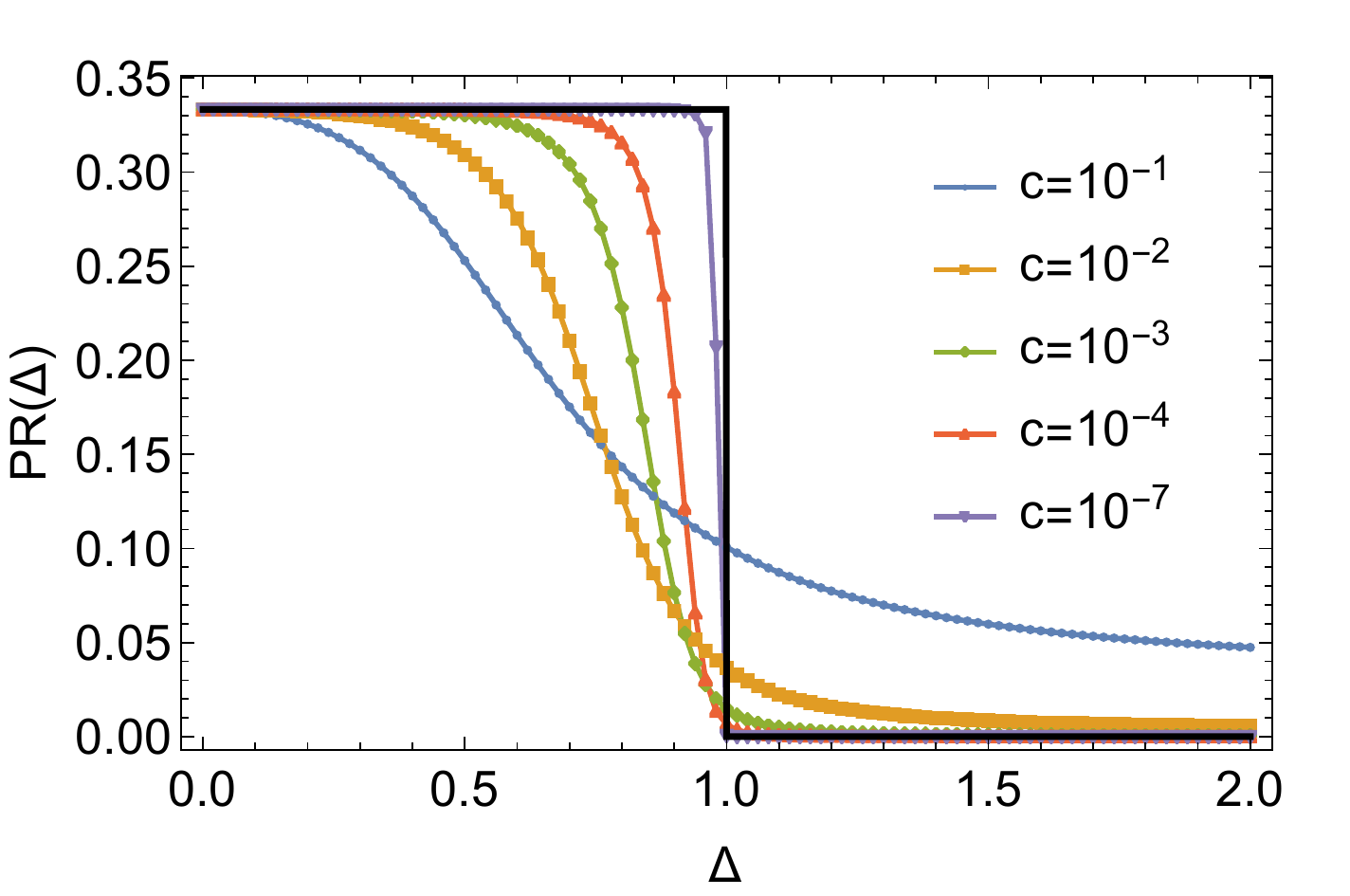} \caption{$\Delta$ dependence of
the participation ratio $\pr$ for the binary distribution. Markers
denote results for $c>0$, while the solid line denotes the result for
$c=0$.}  \label{133942_31Jul22}
\end{center}
\end{figure}
In Fig.~\ref{133942_31Jul22}, we plot $\pr$ for several $c$ to see how the
results converge to Eq.~(\ref{141607_31Jul22}) in the limit $c\to 0$.

\subsection{Continuous distribution: Bose-Einstein condensation}
\begin{figure}[t]
\begin{center}
\includegraphics[width=10cm]{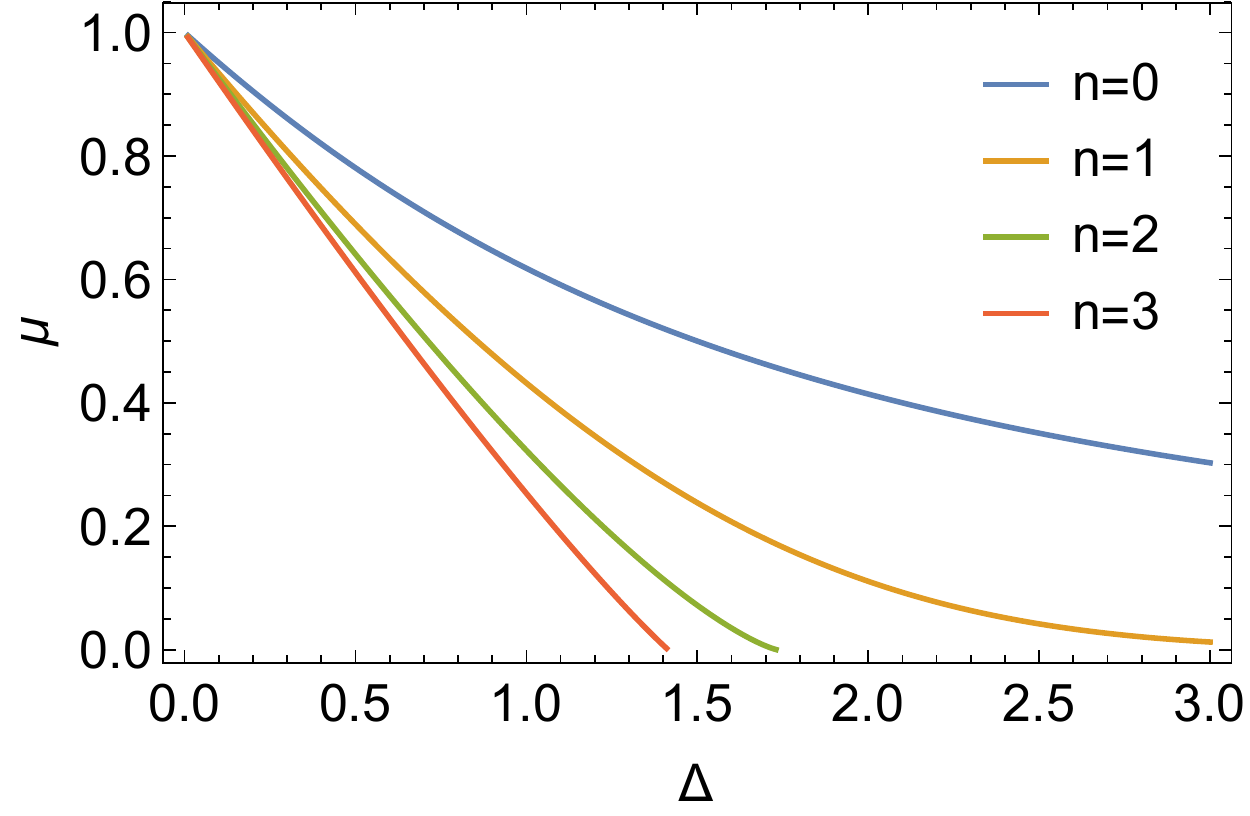} \caption{$\Delta$ dependence of
 the Lagrange multipliyer for the continuous distribution. For $n>1$, we
 plot the data only for $\Delta\leq \Delta_c$.}  \label{085227_1Aug22}
\end{center}
\end{figure}

In the limit $M\to\infty$, one expects that $P(h)$ is approximated by a
continuous function. To simplify the calculation, here we only consider
the following function:
\begin{align}
 P(h) =
\begin{cases}
(1+n)h^n/\Delta^{1+n} & h\in [0,\Delta],\\
0        & {\rm otherwise}
\end{cases},\label{172527_31Jul22}
\end{align}
where $\Delta$ is a positive constant. The pre-factor has been chosen
so that $\int_{-\infty}^\infty dhP(h)=1$. The Lagrange multiplier is
determined by the spherical constraint:
\begin{align}
 1 = \int_{-\infty}^\infty dh P(h)q(h)
 =  \frac{(1+n)}{\Delta^{n+1}}\int_0^\Delta
 \frac{h^{n}dh}{(h+\mu)^2}.\label{170214_31Jul22}
\end{align}
In Fig.~\ref{085227_1Aug22}, we plot the results for several $n$.
The integral in Eq.~(\ref{170214_31Jul22}) takes a maximum at $\mu=0$
($\mu$ can not be negative due to Eq.~(\ref{024515_31Jul22})). If $n>1$, the integral at $\mu=0$ converges
to a finite value :
\begin{align}
\frac{1+n}{\Delta^{n+1}}\int_0^\Delta h^{n-2}dh = \frac{n+1}{\Delta^2(n-1)}.
\end{align}
When $1>\Delta^{-2}(n+1)/(n-1)$ or equivalently
\begin{align}
 \Delta > \Delta_c \equiv \sqrt{\frac{n+1}{n-1}},
\end{align}
Eq.~(\ref{170214_31Jul22}) has no solution. This is similar to the
situation of the previous section, and the term corresponding to $h_k=0$
should be carefully treated. For this purpose, let we explicitly write
down the summation in Eq.~(\ref{203359_30Jul22}) as
\begin{align}
1 = \frac{1}{M}\sum_{k=1}^M q_k=\frac{1}{M}\sum_{k=1}^M\frac{1}{(\mu+h_k)^2},\label{122132_1Aug22}
\end{align}
where 
\begin{align}
 h_k = \Delta \left(\frac{k-1}{M}\right)^{\frac{1}{n+1}}.\label{173833_31Jul22}
\end{align}
Eq.~(\ref{173833_31Jul22}) guarantees that the distribution of $h_k$
converges to Eq.~(\ref{172527_31Jul22}) in the limit $M\to\infty$. A
necessary condition for the sum to be rewritten as an integral is that
each term of the sum goes to zero in the limit of $M\to\infty$.
Below we will check this condition.
The terms for $k>1$ are evaluated as 
\begin{align}
\frac{q_k}{M} = \frac{1}{M(h_k+\mu)^2} = O(M^{-\frac{n-1}{n+1}}),
\end{align}
where we used $h_k=O(M^{-\frac{1}{n+1}})$, see
Eq.~(\ref{173833_31Jul22}). Therefore, $q_k/M\to 0$ if $n>1$. This is not
true for the first term
\begin{align}
 \frac{q_1}{M} = \frac{1}{M\mu^2},
\end{align}
when $\mu\sim 0$. From the above consideration, one realizes that the
first and other terms should be treated separately to rewrite the sum to
an integral for $\Delta>\Delta_c$. In the limit $M\to\infty$, we obtain
\begin{align}
& \frac{q_1}{M}
 + \frac{1}{M}\sum_{k=2}^Mq_k \to
\frac{q_1}{M} + \int_{-\infty}^\infty dhP(h)q(h)
=
 \frac{q_1}{M}
 + \frac{n+1}{\Delta^2(n-1)}.
\end{align}
Substituting back it into Eq.~(\ref{122132_1Aug22}), we get for $\Delta>\Delta_c$
\begin{align}
 \frac{q_1}{M} = 1-  \frac{n+1}{\Delta^2(n-1)},\label{112055_1Aug22}
\end{align}
which is the essentially the same equation as
Eq.~(\ref{105812_31Jul22}).  Eq.~(\ref{112055_1Aug22}) implies that
above $\Delta_c$, the eigenvector tends to condensate to unperturbed
sites for which $h_k=0$. The mathematical structure that causes the
condensation is very similar to that of the Bose-Einstein condensation,
as mentioned in
Refs.~\cite{stanifer2018simple,franz2022delocalization,franz2022linear}.

\begin{figure}[t]
\begin{center}
\includegraphics[width=10cm]{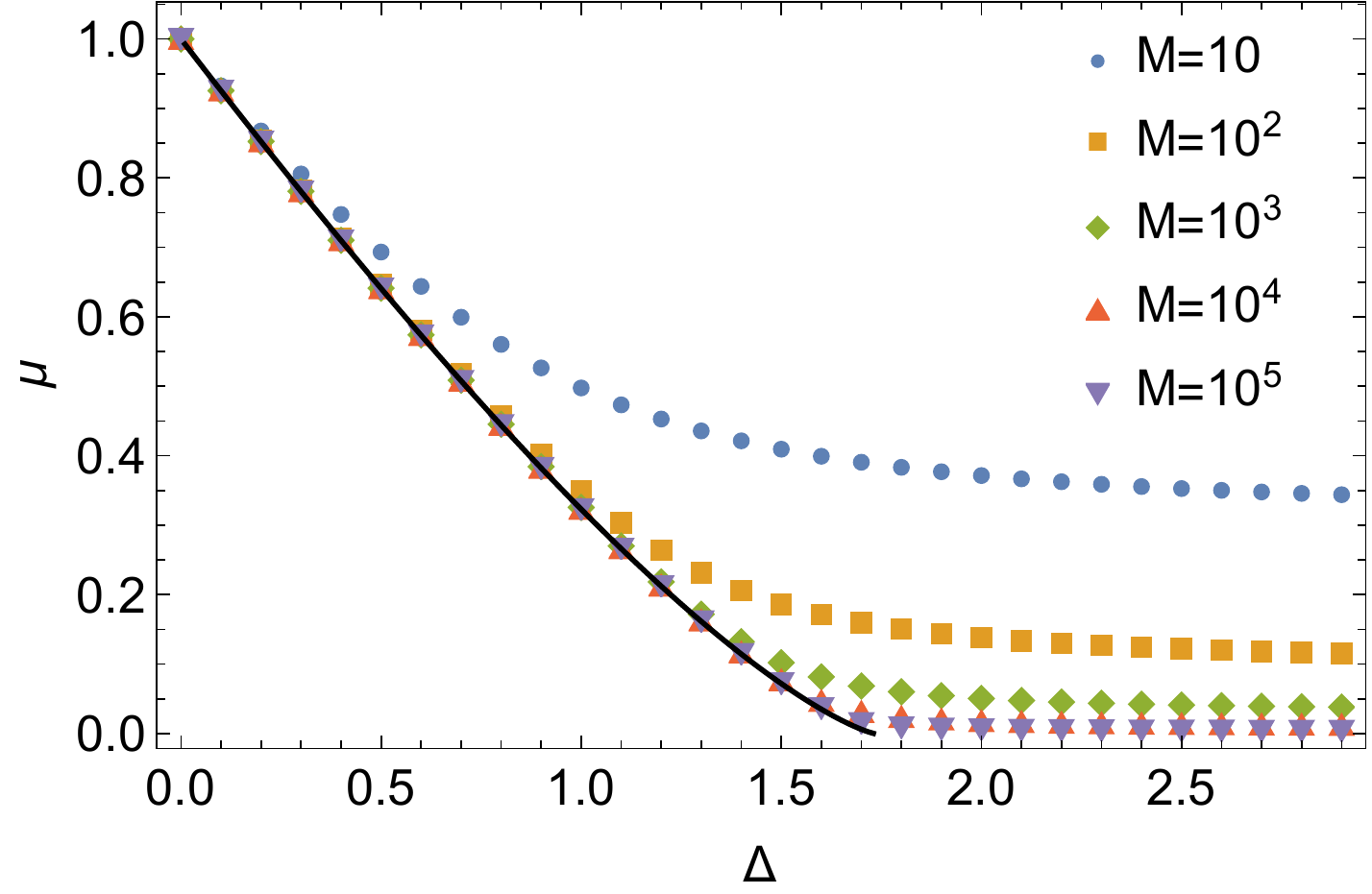} \caption{$\Delta$ dependence of
 the Lagrange multiplier $\mu$ of the continuous distribution for $n=2$
 and for several $M$.  Markers denote results for finite $M$, while the
 solid line denotes the result for $M\to\infty$.}  \label{111930_1Aug22}
\end{center}
\end{figure}

\begin{figure}[t]
\begin{center}
\includegraphics[width=15cm]{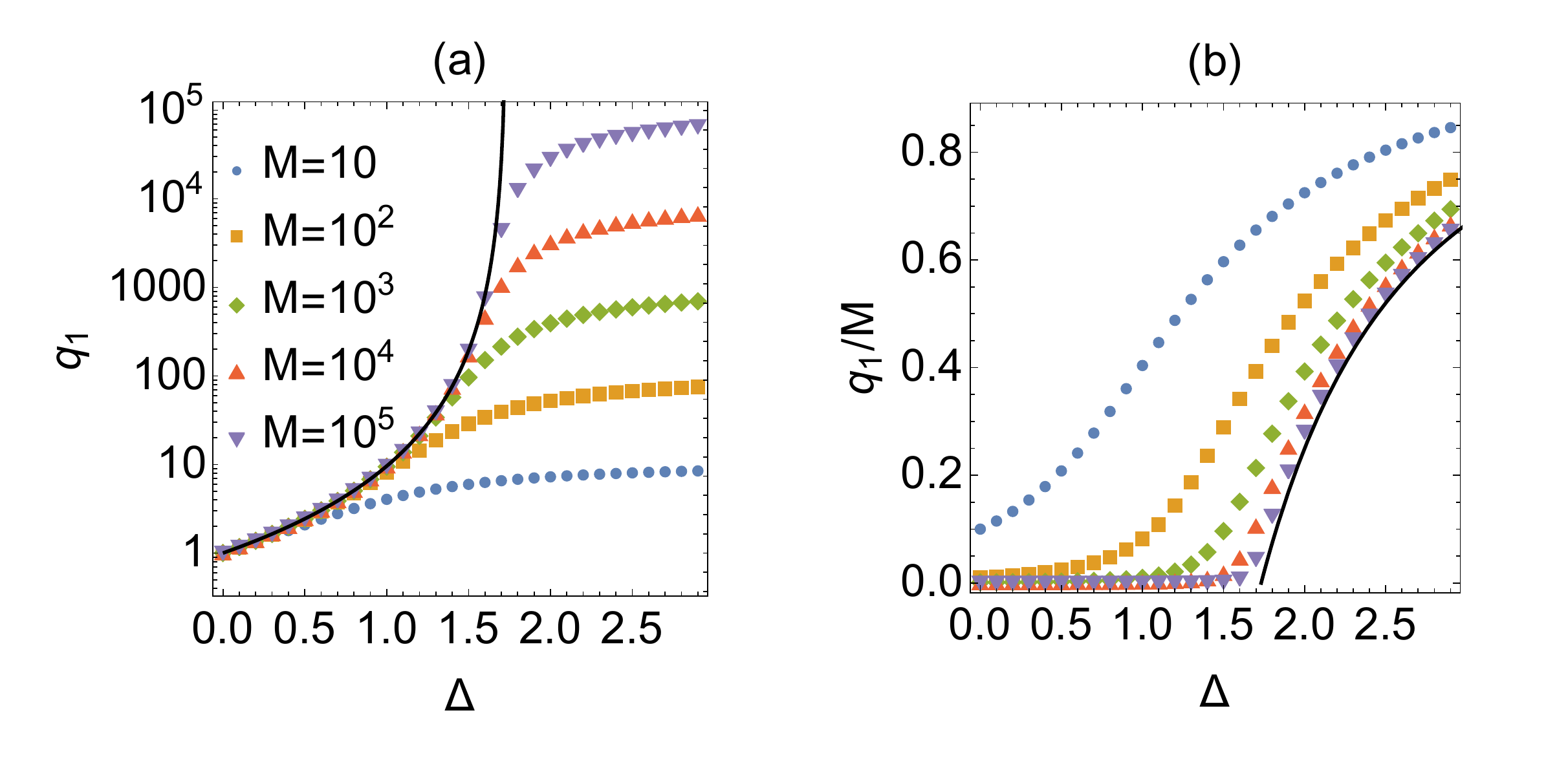} \caption{$\Delta$ dependence of
 the overlap of the continuous distribution for $n=2$. Markers denote results for
 finite $M$, while the solid line denotes the result for $M\to\infty$.}
 \label{111935_1Aug22}
\end{center}
\end{figure}

In Fig.~\ref{111930_1Aug22}, we plot $\mu$ calculated by
Eq.~(\ref{122132_1Aug22}) for $n=2$ and several $M$. For $\Delta\leq
\Delta_c=\sqrt{3}\approx 1.73$, the results nicely converge to that of
the continuum limit $\mu_{\infty}$ calculated by Eq.~(\ref{170214_31Jul22}), while for
$\Delta>\Delta_c$, the results converge to $\mu_{\infty}=0$ in the limit
$M\to\infty$.
In Fig.~\ref{111935_1Aug22}, we plot $q_1$ and $q_1/M$ for several $M$.
For $\Delta\leq \Delta_c$, $q_1$ converges to $1/\mu_\infty^2$ in the
limit $M\to\infty$, see Fig.~\ref{111935_1Aug22}~(a). On the contrary,
for $\Delta>\Delta_c$, $q_1/M$ converges to Eq.~(\ref{112055_1Aug22}),
see Fig.~\ref{111935_1Aug22}~(b).

\begin{figure}[t]
\begin{center}
\includegraphics[width=10cm]{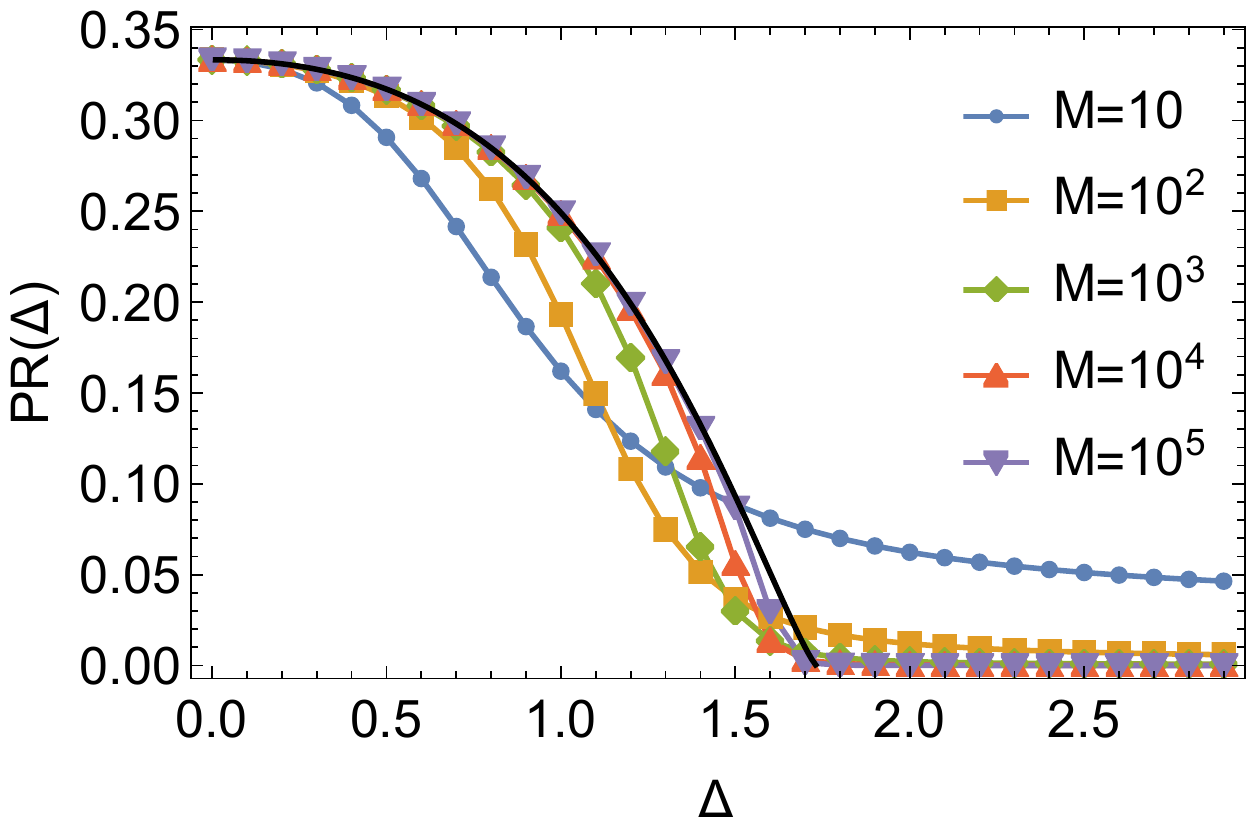} \caption{$\Delta$ dependence of
the participation ratio ${\rm PR}$ of the continuous distribution for
 $n=2$. Markers denote the results for finite $M$, while the solid line
 denotes the result for $M\to\infty$.}  \label{151734_1Aug22}
 \end{center}
\end{figure}

As in Eq.~(\ref{141607_31Jul22}), we use a Gaussian approximation to
calculate the participation ratio~\cite{franz2022delocalization}:
\begin{align}
 \pr = \left[\frac{1}{N}\sum_{i=1}^N\ave{x_i^4}\right]^{-1}\approx 
\left[\frac{1}{N}\sum_{i=1}^N3\ave{x_i}^2\right]^{-1}
= \frac{1}{3}
 \left[\frac{1}{M}\sum_{k=1}^M q_k^2\right]^{-1}.\label{151837_1Aug22}
\end{align}
For $\Delta\leq \Delta_c$, the summation is expressed by an integral, and we get 
\begin{align}
 \pr = \frac{1}{3}\frac{1}{\int_{-\infty}^\infty dh P(h)q(h)^2}.\label{151852_1Aug22}
\end{align}
At the transition point, the denominate is evaluated as 
\begin{align}
 \int dh P(h)q(h)^2 \to \frac{(1+n)}{\Delta^{n+1}}\int_0^\Delta
 h^{n-4}dh = 
\begin{cases}
 \infty & n \leq 3\\
 \frac{n+1}{n-3}\frac{1}{\Delta^2} & n>3
\end{cases}
\end{align}
Therefore, at the transition point, Eq~(\ref{151852_1Aug22}) vanishes
for $n\in(1,3]$ and has a finite value for $n>3$. On the contrary, for
$\Delta>\Delta_c$, the condensation $q_1=O(M)$ leads to $\pr\to 0$ in
the limit $M\to\infty$. Those arguments suggest that on approaching the
transition point, $\pr$ continuously goes to zero for $n\in (1,3]$,
while it changes discontinuously from a finite value to zero for $n>3$.
In Fig.~\ref{151734_1Aug22}, we plot $\pr$ for finite $M$ calculated by
Eq.~(\ref{151837_1Aug22}) and for $M\to\infty$ calculated by
Eq.~(\ref{151852_1Aug22}) for $n=2$. One can see that$\pr$ changes
continuously at $\Delta_c$, in contrast with the binary distribution
where $\pr$ changes discontinuously at the transition point, see
Fig.~\ref{133942_31Jul22}.

\begin{figure}[t]
\begin{center}
\includegraphics[width=13cm]{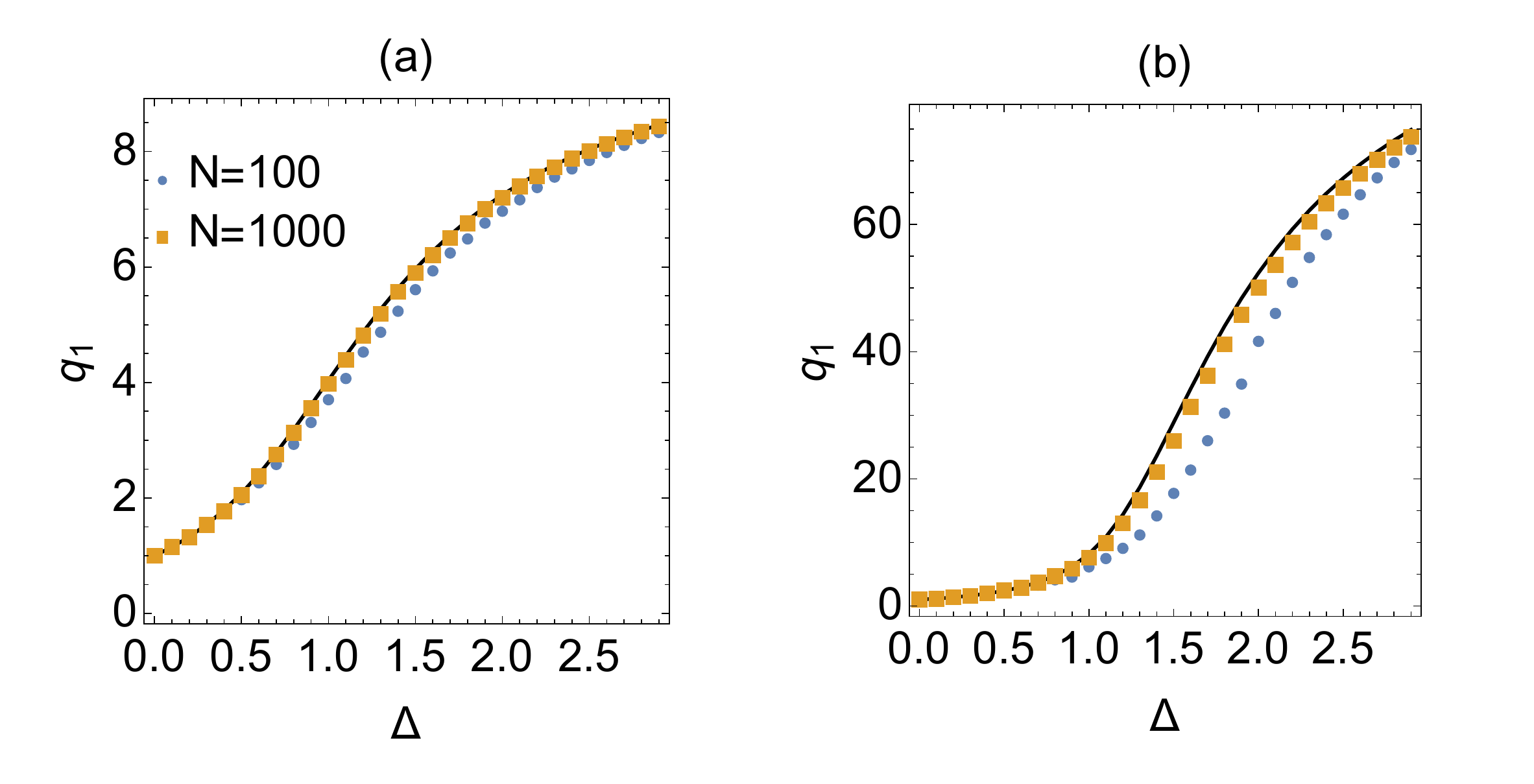} \caption{ $\Delta$ dependence of
$q_1$ for $n=2$. Markers denote numerical results, while the solid line
denotes the theoretical prediction. (a) Results for $M=10$. (b) Results
for $M=100$.}
				       \label{003122_3Aug22}
 \end{center}
\end{figure}

\begin{figure}[t]
\begin{center}
\includegraphics[width=13cm]{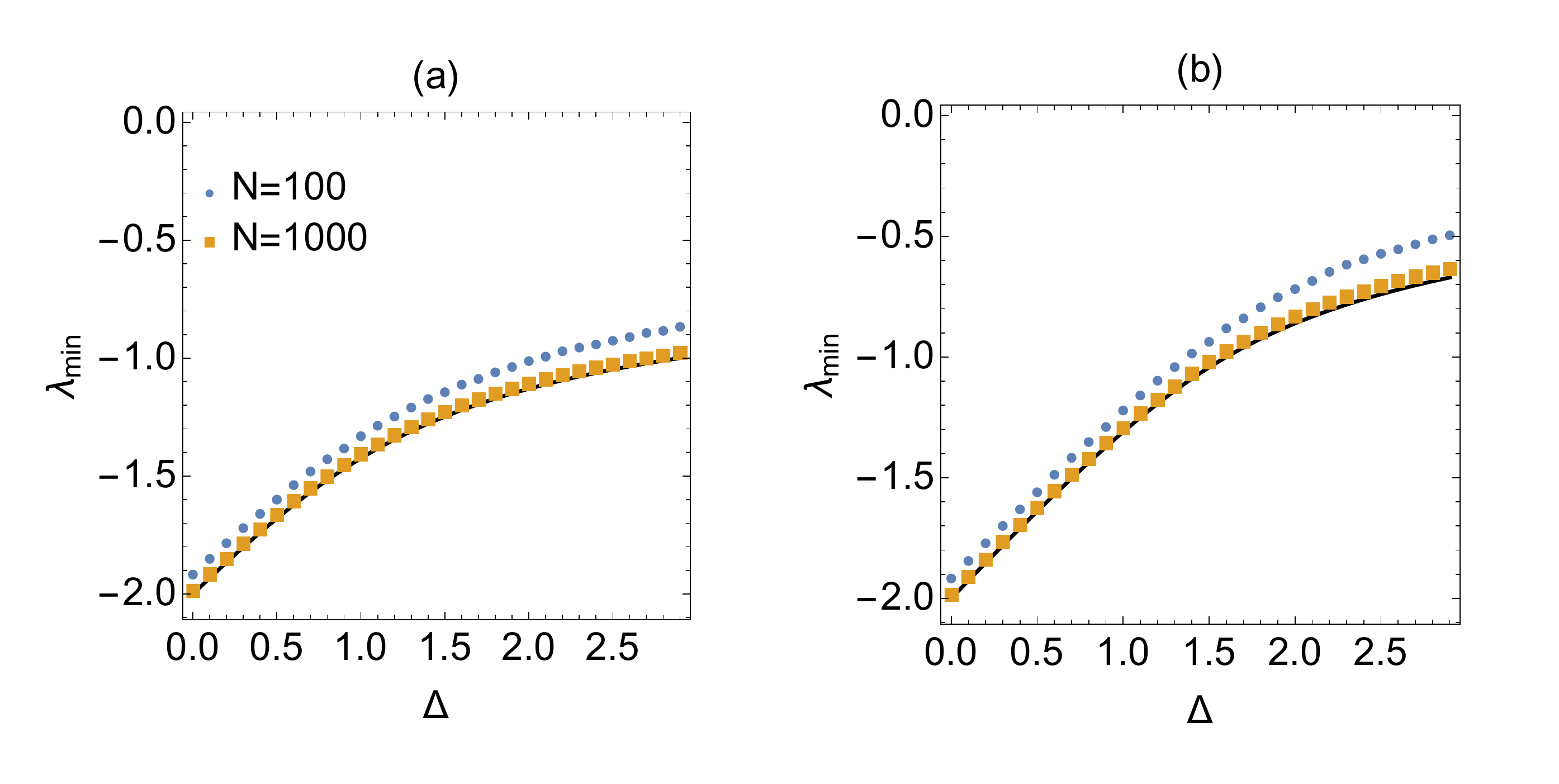} \caption{ $\Delta$ dependence of
$\lambda_{\rm min}$ for $n=2$. Markers denote numerical results, while the solid line
denotes the theoretical prediction. (a) Results for $M=10$. (b) Results
for $M=100$.}
					\label{003126_3Aug22}
 \end{center}
\end{figure}

\begin{figure}[t]
\begin{center}
\includegraphics[width=13cm]{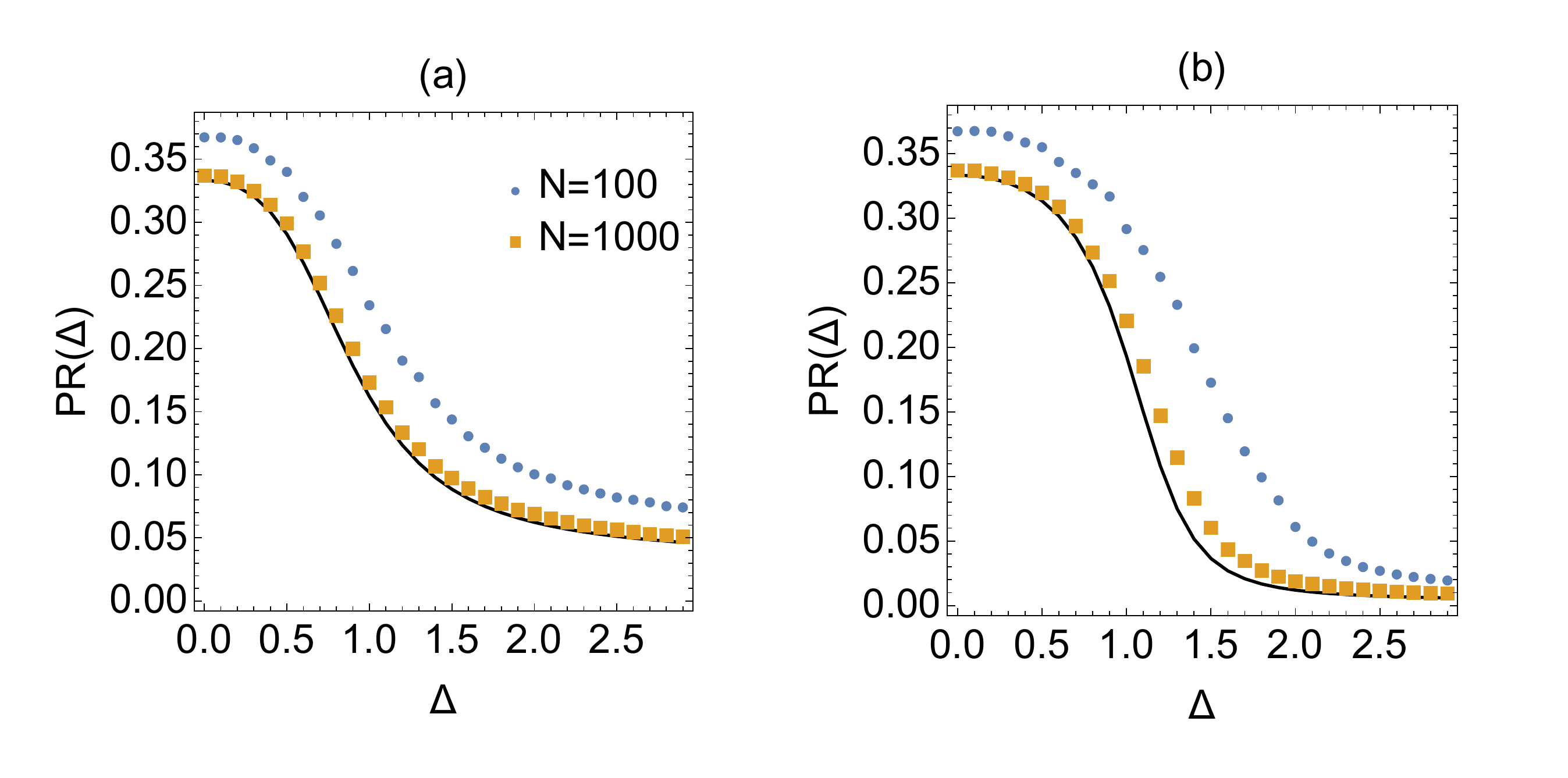} \caption{ $\Delta$ dependence of
$\pr$ for $n=2$. Markers denote numerical results, while the solid line
denotes the theoretical prediction. (a) Results for $M=10$. (b) Results
for $M=100$.}
					\label{003130_3Aug22}
 \end{center}
\end{figure}

Finally, In Figs.~\ref{003122_3Aug22}, \ref{003126_3Aug22}, and
\ref{003130_3Aug22}, we compare the theoretical prediction and numerical
results obtained by direct diagonalization of $W$ for $M=10$ and
$100$. We fond good agreement for $M=10$, while there are small but
visible finite size effects for $M=100$. This is a natural result
because our theory requires $N\gg M$. So we expect larger finite size
effects for larger $M$.

\section{Summary and discussions}
\label{190505_1Aug22}

In this work, we investigated the eigenvector $\bx_{\rm min}$ of the
minimal eigenvalue $\lambda_{\rm min}$ of a deformed random matrix,
where the $i$-th diagonal element of the Wigner matrix is perturbed by a
constant $h_i$. By using the replica method, we closely analyzed the
localization phenomena of $\bx_{\rm min}$ in two cases: when $h_i$ has a
binary distribution, and when it has a continuous distribution.

For the binary distribution of $h_i,$ we considered the following
distribution function: $P(h)=c\delta(h)+(1-c)\delta(h-\Delta)$, where
$c\in [0,1]$ denotes the fraction of non-perturbed sites, and $\Delta>0$
denotes the strength of the perturbation.  On increasing $\Delta$,
$\bx_{\rm min}$ tends to condensate to the non-perturbed sites.  For
$c>0$, this condensation is a crossover: the order parameter just
gradually increases on increasing $\Delta$. As $c$ decreases, the
crossover becomes sharper and eventually becomes a phase transition in
the limit $c\to 0$. At the transition point, the condensation to the
non-perturbed spins leads to a strong localization. As a consequence,
the participation ratio changes discontinuously from a finite value to
zero. In the case of a continuous distribution, we considered a
power-law distribution $P(h)\propto h^n$. We fond that when $n>1$,
$\bx_{\rm min}$ exhibits the Bose-Einstein (like) condensation
transition, as previously fond for a fully-connected vector
spin-glass~\cite{franz2022delocalization}. At transition point,
$\bx_{\rm min}$ tends to condensate to the non-perturbed sites as in the
case of the binary distribution, but this time the participation ratio
goes to zero continuously for $n\in (1,3]$, and discontinuously for
$n>3$.

There are still several important points that deserve further
investigation. Here we give a tentative list:
\begin{itemize}
\item We speculate that the condition $n>1$ for the existence of the
localized phase is somehow universal. Recently, Shimada \textit{et al.}
investigated the localization transition of a $d$-dimensional disordered
lattice by using the effective medium
theory~\cite{shimada2020vibrational,shimada2021novel,shimada2022random}.
They found that for the localized mode to exist, the distribution of the
stiffness $k$ should be $P(k)\sim k^n$ with $n>1$ for $k\ll
1$. Interestingly, this condition is very similar to that we observed in
the case of a continuous distribution of $h_i$. Furthermore, a
phenomenological theory also supports $n>1$~\cite{gurarie2003}. Further
theoretical and numerical studies would be beneficial to clarify this
point~\cite{stanifer2018simple}.

 \item The interaction potential of our model Eq.~(\ref{001213_31Jul22})
is the same of that of the $p=2$-spin spherical model with site
disorders~\cite{castellani2005spin}. In this work, we only investigate
the model at zero temperature. It would be interesting to see how the
model behaves at finite temperatures, which may give some insights for
the thermal excitation of the localized models of amorphous
solids~\cite{das2021,guerra2022}.

\item It is known that for $p>2$, the $p$-spin spherical model exhibits
the one-step replica symmetric breaking
(1RSB)~\cite{castellani2005spin}. Investigating how the 1RSB transition
competes with the condensation transition may provide useful insight
into the competition between glass transition and real-space
condensation~\cite{majumdar2010real}, such as
gelation~\cite{zaccarelli2005,miyazaki2005}.

\item Important future work is to perform a similar calculation for
the Wishart matrix, which has been used to describe the vibrational
density of states of amorphous solids near the jamming transition
point~\cite{franz2015universal}. A recent numerical simulation revealed
that the participation ratio of the lowest localized mode diverges on
approaching the jamming transition point, which characterizes the
correlated volume near the transition point~\cite{shimada2018}.  It may
be possible to derive these behaviors analytically by analyzing a
deformed Wishart matrix.

\item We expect that our method to treat the site randomness can be
      applied to other disordered models.  A promising candidate would
      be the random replicant model (RRM), which is a toy model of
      the coevolution of
      species~\cite{diederich1989replicators,biscari1995replica}. The
      interaction potential of the RRM is written as
      \begin{align}
       H = \sum_{ij}J_{ij}x_i x_j + a\sum_{i=1}^N x_i^2,
      \end{align}
      where $x_i$ denotes the number of the species. The interaction is
      very similar to that of the $p=2$-spin spherical model
      Eq.~(\ref{001213_31Jul22}), but $x_i$ should be positive and
      satisfy the following condition $\sum_{i=1}^N x_i=N$. It is
      interesting to see whether condensation transitions occur when the
      site randomness $\sum_{i}h_ix_i^2$ is added to the RRM, and if so, to investigate
      the implications of the transition for coevolution.
\end{itemize}

\ack  We thank P.~Urbani for useful comments.
This project has received
JSPS KAKENHI Grant Numbers 21K20355.

\appendix 
\section{Binary distribution in the limit $c\to 0$ and Baik-Ben Arous-P\'ech\'e
(BBP) transition} Here we briefly discuss that the transition in the
$c\to 0$ limit of the binary distribution can be identified with the
Baik-Ben Arous-P\'ech\'e (BBP) transition.  A typical setting of the BBP
transition is to add a rank-one perturbation to the Wishart matrix $J$:
\begin{align} 
J+\Delta\be_i\be_i^t,
\end{align}
where $\be_i$ denotes the unit vector along the $i$-th axis. Since the
qualitative results do not depend on $i$, we will set $i=1$ in the following.  The
maximal eigenvalue of the above matrix has been studied extensively, and
it is known that in the thermodynamic limit
$N\to\infty$~\cite{potters2020first}
\begin{align}
 \lambda_{\rm max} = 
\begin{cases}
 2 & \Delta \leq 1 \\
 \Delta + 1/\Delta & \Delta>1.
\end{cases}\label{164940_14Dec22}
\end{align}
The maximal eigenvalue $\lambda_{\rm max}$ exhibits a singular behavior at
the critical point $\Delta_c=1$, which is the signature of the BBP
transition~\cite{potters2020first}.

Now we dicuss that the BBP transition can be identified with the
transition of our model with the binary distribution in the limit $c\to
0$. The matrix $W$ with the binary distribution can be written explicitly
as follows:
\begin{align}
 W = J + \Delta I-\Delta \sum_{i=1}^{cN}\be_i\be_i^t,
\end{align}
where $\be_i$ denotes the unit vector along the $i$-th axis, and $I$
denotes the $N\times N$ identity matrix. 
The minimal eigenvalue is expressed as
\begin{align}
 \lambda_{\rm min}(c) &= \min_{\be}\be^t W\be
 =-\lambda_{\rm max}(c) + \Delta\label{175701_14Dec22} 
\end{align}
where $\be$ denotes an unit vector, and 
\begin{align}
&\lambda_{\rm max}(c)
= -\min_{\be}\be^t\left(J-\Delta\sum_{i=1}^{cN}\be_i\be_i^t\right)\be
= \max_{\be}\be^t\left(J'+\Delta\sum_{i=1}^{cN}\be_i\be_i^t\right)\be,\new
 &J' = -J.\label{164622_14Dec22}
\end{align}
Since the distribution of $J_{ij}$ is symmetric, $J'$ has the same
statistical properties of those of $J$. The question is if $\lambda_{\rm
max}(c)$ converges to the result of the rank-one perturbation
Eq.~(\ref{164940_14Dec22}) in the limit $c\to 0$. The answer is yes: by
substituting Eq.~(\ref{c0}) into (\ref{175701_14Dec22}), one can easily
show that $\lim_{c\to 0}\lambda_{\rm max}(c)=\lambda_{\rm max}$. This
means that the singularity of $\lim_{c\to 0}\lambda_{\rm max}(c)$, or
equivalently $\lim_{c\to 0}\lambda_{\rm min}(c)$, of our model is the
consequence of the BBP transition.

 \section*{References}
\bibliographystyle{iopart-num.bst}
\bibliography{reference}

\end{document}